\documentstyle{mn}
\input{epsf}

\def\fun#1#2{\lower3.6pt\vbox{\baselineskip0pt\lineskip.9pt
\ialign{$\mathsurround=0pt#1\hfill##\hfil$\crcr#2\crcr\sim\crcr}}}

\def\beq{\begin{equation}}
\def\eeq{\end{equation}}
\def\bea{\begin{eqnarray}}
\def\eea{\end{eqnarray}}

\def\mpc {h^{-1} {\rm Mpc}}

\def\LCDM {{$\Lambda$CDM}}

\begin{document}
\title[The 3-point function in large scale structure: redshift distortions and galaxy bias]
{The 3-point function in  large scale structure: \\
redshift distortions and galaxy bias}
\author[ E. Gazta\~naga \& R. Scoccimarro ]
{ E. Gazta\~naga$^{1}$ \& R. Scoccimarro$^{3,4}$ \\
$^{1}$Instituto de Ciencias del Espacio (IEEC/CSIC),
Facultat de Ciencies UAB, Torre C5- Par- 2a, Bellaterra (08193 BARCELONA)  
 \\
$^3$Center for Cosmology and Particle Physics, Department of Physics, New York University, NY, USA
\\
${}^{4}$Kavli Institute for Cosmological Physics, University of Chicago,
Chicago, IL 60637, USA
}
\date{\today}
\pubyear{2004}
\maketitle
\begin{abstract}

We study the behavior of the three-point correlation function  $\zeta$ of dark matter and mock galaxies,  concentrating on the effects of redshift-space distortions and the determination of galaxy bias parameters in current redshift galaxy surveys. On large scales, redshift space distortions tend to wash out slightly the configuration dependence of the reduced 3-point function $Q_3 \sim \zeta/\xi^2$.  On smaller scales ($\le 10 \mpc$), $Q_3$ develops a characteristic U-shape anisotropy between elongated  and open triangles due to the effects of velocity dispersion. We show that this shape is quite universal, very weakly dependent on scale, initial spectral index or cosmological parameters and  should be detectable in current  galaxy surveys  even if affected by shot-noise or galaxy bias. 
We present a detailed method for obtaining constraints on galaxy bias parameters from measurements of $Q_3$ in current galaxy redshift surveys, based on the eigenmode analysis similar to the one developed for the bispectrum.  We show that our method recovers the bias parameters introduced into mock galaxies by a HOD prescription and is also able to handle potential systematics in the case when a smaller number than ideal of mock catalogs is used to estimate the covariance matrix.  We find that current redshift surveys (e.g. SDSS or 2dFGRS) are just about large enough to get interesting new constraints on bias.

\end{abstract}

\section{Introduction}

The galaxy  three-point function provides a valuable statistical tool to investigate the nature of the relationship between galaxies and dark matter and to probe the statistical properties of primordial fluctuations through constraints on primordial non-Gaussianity (see 
Bernardeau et al. 2002 for a review). 

Measurements of the three-point function and other higher-order statistics in galaxy catalogs have a rich history (Peebles \& Groth 1975, Fry  \& Peebles 1978, Baumgart \& Fry 1991, Gazta\~naga 1992, Bouchet et al. 1993, Fry \& Gazta\~naga 1994). In the past decade, three-point statistics have confirmed the basic picture of gravitational instability from Gaussian initial conditions  (Frieman \& Gazta\~naga 1994, Jing \& B\"orner 1998, Frieman \& Gazta\~naga 1999,  Feldman et al. 2001). The impact of such measurements on theoretical models has been important but the systematic and statistical uncertainties in the data have been non-negligible, primarily because the connection between these statistics and theoretical predictions is best done on large scales, where the physics is best understood, and the surveys previously available were not large enough. 

In recent years, with the completion of large redshift surveys such as 2dFGRS (Colless et al. 2001) and SDSS (York et al. 2000) the measurement of higher-order statistics promises to provide tight constraints at the few percent level (Colombi et al. 1998, Szapudi et al. 1999, Matarrese et al 1997, Scoccimarro et al. 2004, Sefusatti \& Scoccimarro 2004) that will help significantly in pinning down the cosmological parameters and issues in galaxy formation that have an impact on galaxy clustering. 
First measurements of three-point statistics in these redshift
surveys are given in Verde et al. (2002), Jing \& B\"orner 2004, Kayo et al. (2004) and Wang etal (2004).

In this paper we study the three-point function with particular emphasis on redshift-space distortions and galaxy bias. There is a puzzling result regarding previous investigations of the 3-point function in redshift
(configuration) space (Suto \& Matsubara 1994, Matsubara \& Suto 1994, Jing \& B\"orner  1998,
Jing \& B\"orner 2004).  They all find  very little dependence of the 3-point function on triangle configuration at small scales (both in simulations and observations). This result is puzzling because one of the first things that brings attention to our eyes is the presence of the ``fingers of god" in redshift-space distributions, produced by the radial component of  galaxy peculiar velocities inside clusters.  These rather long and thin structures should leave a clear  imprint in the 3-point correlation function, with higher amplitudes for collapsed triangles. Why has this feature not been observed so far? 

We also study the three-point function as a probe of galaxy bias in the framework of current surveys, and provide a detailed framework for computing reliable error estimates on bias parameters that include the covariance matrix due to non-linearities and survey geometry. We use a battery of N-body simulations and mock catalogs extracted from them with different biasing prescriptions. Our presentation follows closely the eigenmode analysis method developed for the bispectrum (Scoccimarro 2000). 

A first paper of this series (Barriga \& Gazta\~naga 2002, paper I) presented a comparison of the predictions for the two and three-point correlation functions of density fluctuations, $\xi$ and $\zeta$, in
gravitational perturbation theory (PT) against large Cold Dark Matter (CDM) simulations. Here we extend these results into the non-linear regime and focus on the effects of redshift distortions and the extraction of galaxy bias parameters in galaxy surveys. This paper is organized as follows. In section~\ref{3pt} we review the basic definitions involving the three-point function. In section~\ref{simu} we present the simulations and mock catalogs, whereas in section~\ref{zdist} we study the effects of redshift distortions from large to small scales. In section~\ref{getbias} we study in detail how to recover bias parameters from current galaxy catalogs. 

\section{The three-point function}
\label{3pt}

\subsection{Definitions}

The two and three-point correlation functions are defined, respectively, as
\bea
\xi(r_{12}) &=& \langle \delta(r_1) \delta(r_2) \rangle \\
\zeta(r_{12},r_{23},r_{13}) &=& \langle \delta(r_1) \delta(r_2) \delta(r_3) \rangle 
\eea
where $\delta(r) = \rho(r)/\bar{\rho}-1$ is the local density fluctuation about 
the mean $\bar{\rho}=\langle\rho\rangle$, and the expectation value is take over different
realizations of the model or physical process. In practice, the expectation
value is over different spatial positions in our Universe, which are
assumed to be a fair sample of possible realizations (see Peebles 1980). It is convenient to define a 
$Q_3$ parameter as  follows~\cite{GP77}
\begin{eqnarray}
Q_3 &=& \frac{\zeta(r_{12},r_{23},r_{13})}{\zeta_H(r_{12},r_{23},r_{13})} \\
\zeta_H  &\equiv& 
{\xi (r_{12})\xi(r_{23})+\xi(\vec r_{12})\xi(r_{13})+\xi(r_{23})\xi(r_{13})},
\label{fiftheq}
\end{eqnarray}
\noindent
where we have introduced a definition for the "hierarchical" three-point function $\zeta_H$.
The $Q_3$ parameter was thought to be roughly constant 
as a function of triangle shape and scale~\cite{PB80}, a result that is usually 
referred to as the hierarchical scaling. As we will see below, accurate 
measurements/predictions show that $Q_3$ is not quite constant in any regime of
clustering, although the variations of $Q_3$ with scale are small compared to the
corresponding  changes in $\xi$ or $\zeta$. 

\subsection{Triangle parametrizations:  $\alpha$ vs. $v$}
\label{sec:coll}

There are two parametrizations that are commonly used in the literature to describe the three parameters that define a triangle, each of which has advantages and disadvantages. Since we are studying the monopole of the three-point function (averaged over all orientations of the triangle with respect to the observer), this depends on three variables. One obvious choice is the magnitude of the three sides of the triangle, described by  $\vec r_{12}$, $\vec r_{23}$ and $\vec r_{31}= \vec r_{12} + \vec r_{23}$. This is reasonable since the three variables are on an equal footing (being three lengths) and if the measurement is done with some typical resolution due to the three-point function algorithm, it affects all of them in the same way. 

The two most common parametrizations, however, involve a length scale, a dimensionless number (ratio of lengths), and some measure of triangle shape (another ratio of lengths or an angle). In this case the three variables may be affected by resolution (e.g. binning) in different ways. This can be important for example for the dependence on triangle shape, which is what concerns us here regarding the signature of fingers of God at small scales.

In perturbation theory,
it has been customary to use the two sides
of the triangle $r_{12}$ and $r_{23}$ (which are typically
comparable in side) and the angle 
$\alpha$ between them:

\beq
cos(\alpha)= {\vec r_{12}\over{r_{12}}} . {\vec r_{23}\over{r_{23}}}
\eeq
with $0<\alpha<180 $ deg. For small scales it has been popular to use $r_{12}$ and $r_{23}$,
but with later  given by the ratio:
\beq
u= r_{23}/r_{12},
\eeq
and the third parameter given by:
\beq
v \equiv {r_{31}-r_{23} \over{r_{12}}} 
\eeq
with the restriction $r_{12}<r_{23}<r_{13}$.
The value of $v$ is related to $\alpha$ by:
\beq
v = \sqrt{u^2 - 2 ~u~\cos \alpha +1} - u
\eeq
In principle $-1<v<1$, but the restriction 
$r_{12}<r_{23}<r_{13}$ only allows $0<v<1$. 
The minimum value of $v=0$ corresponds to $\cos \alpha= 1/(2u)$.
Note how in this parametrization the small values of $\alpha$
are only recovered for large values of $u$. For example,
for $u=1$ we are restricted to $60 < \alpha < 180$ deg. This can have implications for the detection of sharp dependencies on triangle shape, as we shall see in 
section~\ref{zdist}.


Elongated or ``collapsed configurations" are those with $\alpha \simeq 0$ or
$\alpha \simeq 180$ deg; or in terms of $v$, those with $v$ close
to unity. Configurations of triangles with $\alpha \simeq 90$
are called here ``perpendicular" configurations.   In terms of 
$v$ they correspond to small values of $v$, but note that there is
no one-to-one correspondence. Small values of $v$ could also
arise for triangles with $r_{31}=r_{23}$ even if $r_{12}$ is not
small. 

We  use the term "strong configuration dependence"  when there is a significance
difference between the collapsed and the perpendicular configurations. By 
"weak configuration dependence" we mean that $Q_3$ is hierarchical
 (ie constant as a function of $\alpha$).

\begin{table*}
\begin{tabular}{|c|c|c|c|c|c|c|clclc|} \hline
                & \LCDM400&  SCDM400 & \LCDM  & $\tau CDM$ &  $OCDM$ &  $SCDM$ & VLS & GIF & HV \\ \hline
\hline
$\Omega _m$        & 0.3	& 1.0 	&  0.3     	&  1.0      	&  0.3		& 1.0  & 0.3 & 0.3 & 0.3\\ \hline
$\Omega _{\Lambda}$ &	0.7 &	0.0 	&  0.7     	&  0.0      	&  0.0		& 0.0  & 0.7 & 0.7 & 0.7\\ \hline
$h$                &  0.7	 &  0.5	& 0.7     	&  0.7		    &  0.7 		& 0.5 & 0.7 & 0.7 & 0.7\\ \hline
$\Gamma $          &  0.2   &  0.5		& 0.2     	&  0.2		    & 	0.2		& 0.5 & 0.16 & 0.21 & 0.16 \\ \hline
$\sigma_8$		   & 1.0 & 1.0		& 0.9		&  0.51			& 	0.85		& 0.51& 0.9& 0.9 & 0.9\\ \hline
$N_{\rm par}$            &  $126^3$   &  $126^3$ 	& $256^3$ 	& $256^3$ 		& $256^3$  & $256^3$  & $512^3$& $256^3$ & $1000^3 $\\ \hline
$L_{\rm size}$       & $400$ & $400$	&$240$ & $240$ 	& $240$	 & $240$ & $479$ & $141$ & $3000$\\  \hline
\end{tabular}
\caption[junk]{Main parameters describing the simulations. Box size $L_{\rm size}$ is in $\mpc$. 
For VLS and HV the shape parameter $\Gamma$ stands for an effective value corresponding to the linear power spectrum with a transfer function for $\Omega_m=0.26$, $\Omega_b=0.04$ and $\Omega_{\Lambda}=0.7=h$.}
\label{tab1}
\end{table*}

\subsection{3-point Function Estimator and Algorithm}

We follow the fast algorithm described in some detail in paper I.
This algorithm allows a fast calculation of two and three-point function for millions
of points. 
The first step is to discretize the simulation box
into $Lsize^3$ cubic cells. We assign each particle to a node of this new
latticed box using the nearest grid point particle assignment.  
We precalculate the list of relative neighbors to any given 
node in the lattice. To compute now the two-point and three-point correlation 
functions we use:
\bea
\xi(r_{12}) &=& 
{\sum_{i,j} \delta_i \delta_j \over{ \sum_{i,j} 1}}
\\
\zeta(r_{12},r_{23},r_{13}) &=&  {\sum_{i,j,k} \delta_i \delta_j \delta_k 
\over{ \sum_{i,j,k} 1}}
\eea
where $i$ extends over all nodes in the lattice, $j$ over the list of
precalculated neighbors (see paper I)
that are at a distance $r_{12}$ from $i$ and
$k$ over the neighbors at distance $r_{23}$ from $j$ and
$r_{13}$ from $i$.


Besides the tests presented in Barriga \& Gazta\~naga (2002) we
have performed a series of additional tests  designed to quantify how robust  our new findings
are in redshift space and on smaller scales.  We have checked that very
similar results are obtained for different pixel sizes. Pixels
can be as large as $r_{12}/3$ on a side with negligible effects
in the resulting $Q_3(\alpha)$. This of course imposes a constraint on how
thick we can take the binning in $\alpha$ and works as long as
$Q_3$ has a smooth shape as a function of $\alpha$. As we will show
below (ie Fig.\ref{q3bzr1}) the pixel scale or the binning play an important role
on small scales.

Dilution of point density by factors of 1/10 or 1/20 
does not bias the results, indicating that the above estimator
is robust. Using realistic mock catalogues (eg. with a 2dFGRS or SDSS mask) instead of cubical boxes  gives very similar results. In addition, the LCDM240 and LCDM400 simulations (see table~\ref{tab1} below) give values of $Q_3$ in excellent agreement at
small scales. 

\section{Simulations}
\label{simu}

\begin{figure*}
\centerline{\epsfysize=8truecm
\epsfbox{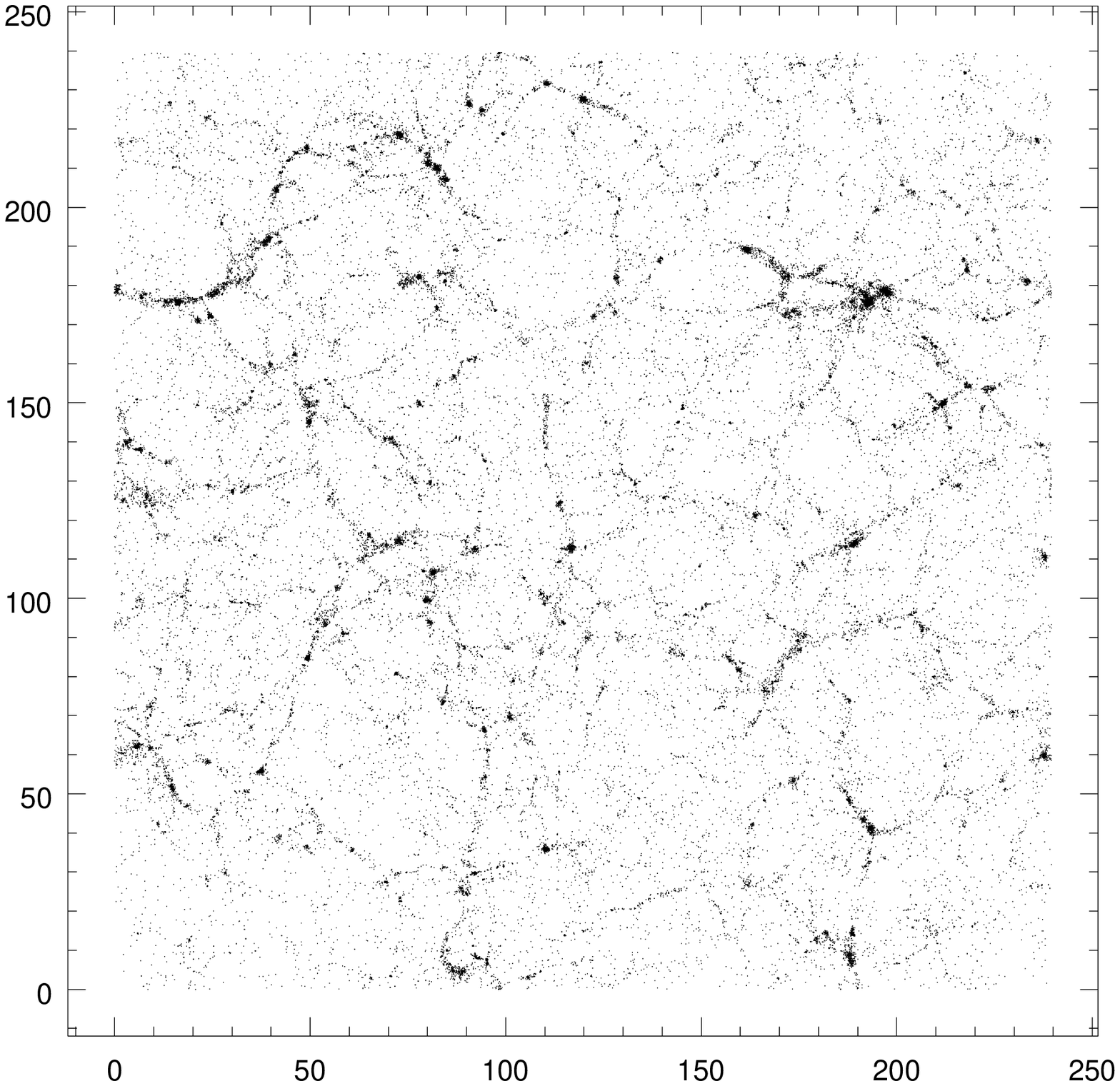}\epsfysize=8truecm
\epsfbox{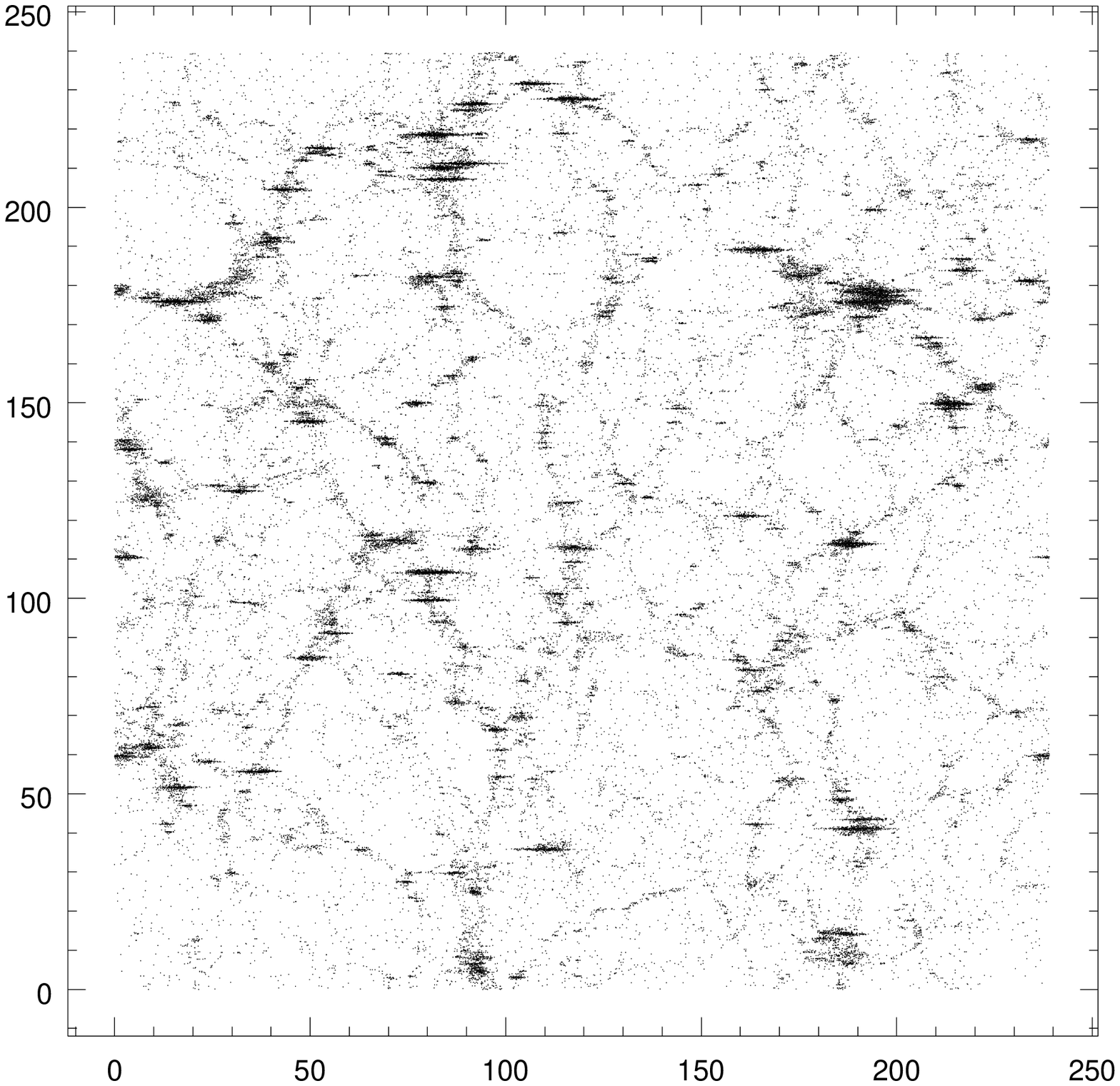}}
\caption[junk]{Particle positions in real (left)
and redshift space (right) in a 1 Mpc/h thick  projected slice from to the LCDM240 simulation (see table~\protect\ref{tab1}).}
\label{plotslice}
\end{figure*}

We use several sets of simulations centered around
 the Lambda Cold Dark Matter model
(\LCDM$~$ from now on) with $\Omega_\Lambda=0.7$, 
$\Omega_m=0.3$, $h=0.7$. 
The first set, which we call \LCDM400, consist of  5 independent 
realizations of  \LCDM$~$ model normalized to $\sigma_8=1$ in 
a box $L=378 \mpc$ with $N=126^3$ particles. These simulations
were run with the AP3M code of Efstathiou et al. (1985) 
to study volume-averaged higher-order correlations
(see Baugh \& Gazta\~naga 1996 for more details). 
The particle mass is $M \simeq 2.3 \times 10^{12} M_{\sun}/h$.

A second set, which we just
call \LCDM240$~$ consists on  a single
realization of the \LCDM$~$ model normalized to $\sigma_8=0.9$ in 
a box $L=239.5 \mpc$ with $N=256^3$ particles. This simulation
was run with a  parallel version  of the AP3M code by the VIRGO consortium.
The particle mass is $M \simeq 6.9 \times 10^{10} M_{sun}/h$.
Thus, while a single \LCDM400 has $\simeq 4$ times more volume
(20 times more volume if we use all 5 realizations) the \LCDM240
model has $\simeq 33$ times higher particle density (and therefore
33 times higher mass resolution). This will allow us to explore how
$Q_3$ is affected by resolution and sampling effects.
The difference in normalization ($\sigma_8=1.0$ compared to 
 $\sigma_8=0.9$) is quite small and has very little impact on
$Q_3$.

We also study three further different models using simulations from the VIRGO consortium 
with identical volume and number of particles as \LCDM240. The different
models are listed in Table~\ref{tab1}. 
The comparison with the SCDM allows us 
to study the dependence of $Q_3$ on the shape parameter $\Gamma$, while the rest of the models explore sensitivity to cosmological parameters and normalization. 

Finally, we consider two other simulations from the VIRGO consortium, the GIF (higher resolution),
 VLS (larger volume) and HV (even larger volume but with lower resolution) simulations, also shown
in  Table~\ref{tab1}. These in particular were used to build the galaxy mock catalogs  by using the Halo Occupation Distribution (HOD) formalism (Ma and Fry 2000, Peacock and Smith 2000, Seljak 2000, Scoccimarro et al. 2001, Berlind et al 2003, see Cooray and Sheth 2002 for a review), see Section~\ref{mocks} for details.


\begin{figure*}
\centerline{\epsfysize=12truecm
\epsfbox{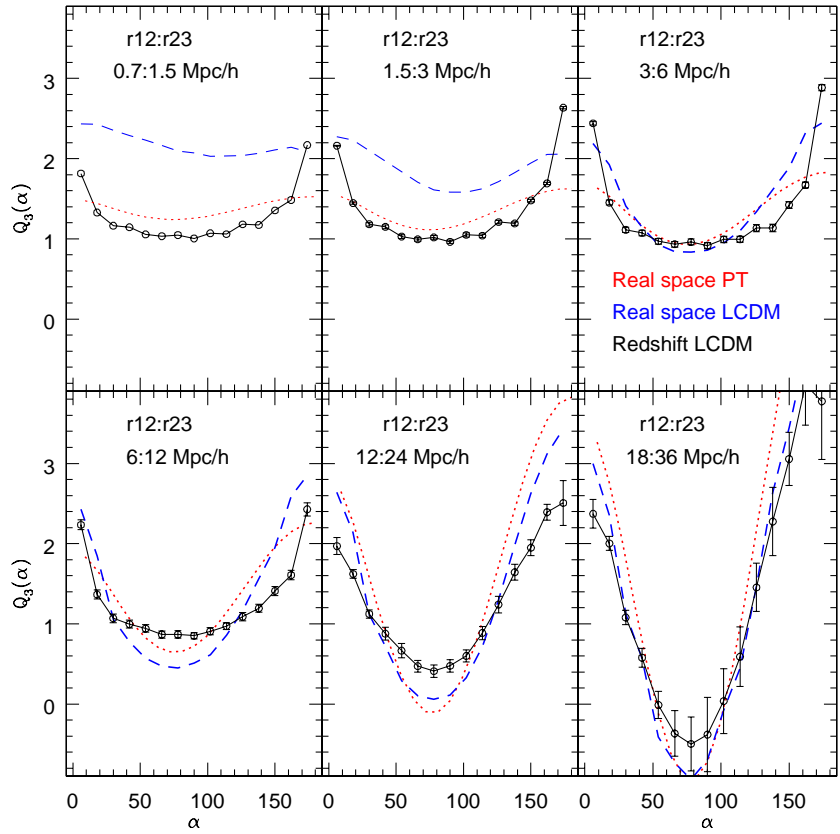}}
\caption[junk]{$Q_3(r_{12},r_{23},\alpha)$ 
 as a  function $\alpha$,  the angle between $\vec r_{12}$ and 
$\vec r_{23}$, in \LCDM$~$400 simulations.  
Real space measurements are shown as dashed lines,  while dotted lines show 
tree-level (real-space) perturbation theory predictions. 
Symbols with errorbars correspond to redshift space measurements.
Each panel shows the results for triangles of
different scales.}
\label{q3zc80r12}
\end{figure*}

\section{$Q_3$ in redshift space}
\label{zdist}

Redshift distortions (Jackson 1972, Sargent and Turner 1977, Peebles 1980, Kaiser 1987) due to peculiar velocities along the line of sight create an anisotropic galaxy distribution with a preferred direction, that joining the observer to the each galaxy. Therefore different survey geometries will
be affected differently by redshift distortions. Here we focus on  the long-distance observer approximation
where we place the observer in one of the axes at infinity. This allows a general study
of redshift distortions, independent of the survey geometry. Nevertheless we do not
find strong differences for more realistic situations, including that of 
a redshift slice (where one of the angular distances, typically declination,
is restricted to a few degrees on the sky). 

Figure~\ref{plotslice} shows a 1 Mpc/$h$   slice in the \LCDM240 case. One can see that:

a) fingers of God are long and sharp needles of about $1 \mpc$ wide
and $10 \mpc$ long. This should make a very noticeable imprint on $Q_3$ for elongated 
triangles (i.e. the "collapsed" configurations in \S\ref{sec:coll}) across those scales. 

b) filamentary structure is apparent on large scales in both cases,
although in redshift space it seems to be more washed away. This should show as
weaker configuration dependence (or less contrast between collapsed and prependicular
triangles) for $Q_3$ in redshift space than in real space at these large scales.

c) on smaller (non-linear) scales structures are rounder, and seem sharper
in real space. Thus, weaker configuration dependence and larger
values of $Q_3$ are expected on small scales in real space.

Let's check how these visual impressions are reflected in the actual measurements
of $Q_3$.

Figure~\ref{q3zc80r12} compares the estimations of $Q_3$ in the 
\LCDM400  case  for  triangles with $| \vec r_{12}|=|\vec
r_{23}|/2=1.5,3,6,9,12,18 \mpc$ in real space (dashed lines) 
and redshift space (symbols with errorbars). The symbols 
with error-bars show the mean and variance in 5 realizations
of the same \LCDM$~$ simulations, while the dotted 
lines show the corresponding leading order (tree-level) 
PT prediction in real space.

On large scales $Q_3$ in redshift space has only slightly less
configuration (or shape) dependence than in real space, as expected from perturbation theory (Hivon et al. 1995) and in agreement with bispectrum results (Scoccimarro et al. 1999),  indicating that part of the angular asymmetries are washed out by the peculiar velocities, but this effect is small.
Note that on the largest scales $18-36 \mpc$, the values of
$Q_3$ both in real (or redshift space) seem to approach the PT
predictions, within the large errorbars. This provides  
an important observational test for the current picture
of structure formation, independent of cosmological parameters for a fixed power spectrum shape 
(see Bernardeau et al. 2002). 

On scales smaller than $\le 10 \mpc$ (where $\xi>1$),
 the redshift space results are almost independent of   the overall scale of the triangle
(despite the large changes in the amplitude of $\xi$),
 while the real space values increase sharply as a function
 of scale (at least up to $\sim 1 \mpc$). 
 Note how $Q_3$ for collapsed configurations $\alpha \simeq 0,180$ deg in redshift space rises
sharply, while for the perpendicular configurations ($\alpha \sim 90$)
the values of $Q_3$ are quite flat and scale independent. This characteristic anisotropy is what we call the {\em  U-shape}. This U-shape does not appear in real space or in PT theory, as it is
basically the results of small-scale redshift distortions, the effect of fingers of God. A similar effect is seen in the small-scale redshift-space bispectrum (see Figs.~3 and~4 in Scoccimarro et al. 1999). 
\begin{figure}
\centering{
{\epsfxsize=8cm \epsfbox{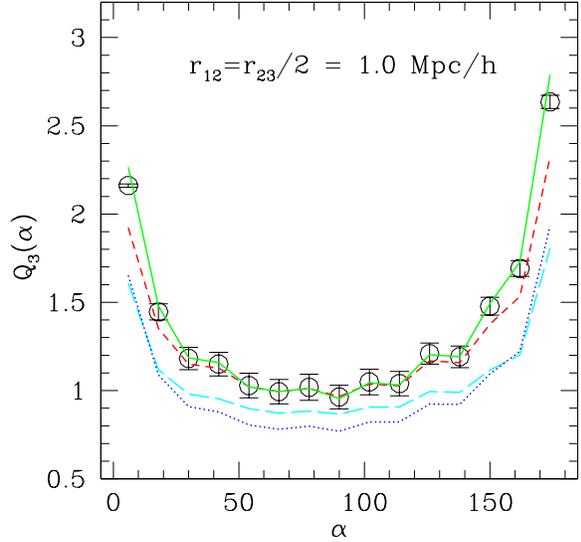}}}
\caption[junk]{$Q_3(r_{12},r_{23},\alpha)$ in redshift space 
 as a  function $\alpha$,  the angle between $\vec r_{12}$ and 
$\vec r_{23}$,  for  simulations in Table \ref{tab1}.  
Symbols with 3-sigma errorbars (from the scatter in 5 realizations)
correspond to \LCDM400. Continuous, short dashed, long dashed and dotted
 lines correspond to OCDM, \LCDM240,  tCDM and SCDM respectively.}
\label{q3zr1}
\end{figure}

\begin{figure}
\centering{
{\epsfxsize=9cm \epsfbox{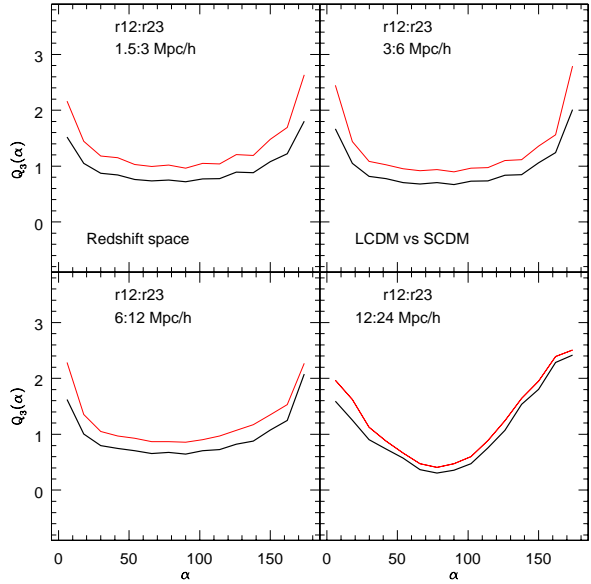}}}
\caption[junk]{Comparison of redshift-space $Q_3(r_{12},r_{23},\alpha)$ for 
SCDM400 (thick lines) and  \LCDM400 (thin lines).}  
\label{q3zc189c80}
\end{figure}

In Fig.~\ref{q3zr1} we compare the values of $Q_3$ in redshift space
for different models on small scales. 
 The U-shape anisotropy is present in all models
regardless of differences in the cosmological parameters. Note also the small errorbars and
 the  good agreement  between the two \LCDM~ models with very different resolution (and slightly different $\sigma_8$ normalization). These results extend to smaller scales: we find very
 similar results (with no evidence for any change with scale) in the range $r_{12}=0.1-3$ Mpc/h.
 All cases in Fig.~\ref{q3zr1} correspond to $r_{12}=r_{23}/2= 1.0$Mpc/h,
except for  the  \LCDM400 (symbols with errorbars) which has  $r_{12}=r_{23}/2= 1.5$Mpc/h. 
 Figure~\ref{q3zc189c80} shows the comparison between \LCDM~ and SCDM
on different scales. As shown in the figure there is a small global 
shift between the two models on all scales. On the largest scales, the
difference become smaller (in contrast to what happens in real
space, see paper I).

The results of Fig.~\ref{q3zr1} show that the U-shape anisotropy is expected in a wide range of models, and as we will see below (see  \S\ref{mocks})
it is also expected in biased galaxy distributions. However, it has not been seen previously  in N-body simulations or observations.  Early measurements in simulations (Suto \& Matsubara 1994,
 Matsubara \& Suto 1994) claimed that the hierarchical ansatz ($Q_3$ constant independent of configuration and scale) is valid in redshift space at small scales. More recent studies (Jing \& B\"orner 1998) see a very mild effect, but nothing comparable to the amplitude seen in Fig.~\ref{q3zr1}. In observations, similar results are seen in LCRS (Jing \& B\"orner 1998), 2dFGRS (Jing \& B\"orner 2004) and SDSS (Kayo et al. 2004). We believe that these discrepancies with our results are due to these authors'  choice of binning, as we discuss now. For example, Matsubara \& Suto (1994) bin distances which vary by $30\%$, Jing \& B\"orner (1998,2004) have only 5 bins in the $v$ variable, comparable to Kayo et al (2004), who use 5 bins in $\alpha$ from 0 to 180 deg.

\begin{figure}
\centering{
{\epsfxsize=8cm \epsfbox{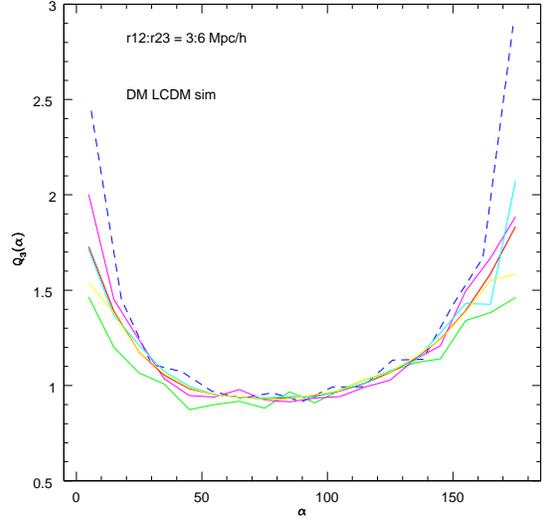}}}
\caption[junk]{Effects of resolution and sparse sampling. The dashed black line shows the result with highest resolution, which gives the higher results for the collapsed configurations ($\alpha=0,180$) .  The lowest line corresponds to the lowest resolution (cubic cells of 1 Mpc/$h$ and sparse sampling of 1 in10 triangles). Intermediate lines shows an increase in resolution by a factor of two (with full sampling).}
\label{q3bzr1}
\end{figure}

In order to investigate possible explanations for this situation,  we explore the effect of using different pixel resolutions and sparse sampling of triangles. We see in Fig.~\ref{q3bzr1} that this typically results in a biased-low value of $Q_3$, significantly so at collapsed configurations. Because there are far fewer triangles in the collapsed configuration, low-resolution translates into a reduced U-shape, effectively mixing a small number of large amplitudes with  large numbers of small amplitudes and thus down-weighting the resulting $Q_3$ at collapsed configurations. A similar effect is obtained if the binning in angle is made too wide; the effect is even more severe in the $v$-variable parametrization (see section~\ref{sec:coll}) which has been traditionally used at small scales. We believe these reasons may explain why the U-shape anisotropy has not been seen before in the three-point correlation function studies.

\section{GALAXY BIASING}
\label{getbias}

We now check whether biasing affects significantly the characteristic
U-shape dependence we find in redshift space at small scales. We will also study what
constraints we can get on biasing parameters from the measurements of $Q_3$.
 
The value of $Q_3$ depends on galaxy bias, in a form that can be easily calculated on large enough scales, where galaxy fluctuations $\delta_G$ can be modeled as a local (but non-linear) function of 
the corresponding matter fluctuations $\delta$;  thus expanding this local function for $\delta\ll 1$ it follows that (Fry \& Gazta\~naga 1993) 
\beq
\delta_G \simeq F[\delta] \simeq \sum_{k}~{b_k\over{k!}}~\delta^k ,
\label{eq:bk}
\eeq
where $k=0$ comes from the requirement that $\langle \delta_G \rangle =0$. It then
follows (see Fry \& Gazta\~naga 1993, Frieman \& Gazta\~naga 1994) that:
\beq
Q_3^G \simeq {1\over{b_1}} ~\left(Q_3+c_2 \right)
\label{eq:Q3G}
\eeq
where $c_2 \equiv b_2/b_1$, and the $\simeq$ sign indicates that this
is the leading order contribution in the expansion given by Eq.~(\ref{eq:bk}) above.  

Thus, in general, the linear bias prescription is not accurate for higher-order moments 
even when $\delta \ll 1$, the reason being that nonlinearities generate non-Gaussianities of the same order as those of gravitational origin. The linear bias term $b_1$ can produce distortions in the shape of $Q_3$, while the non-linear terms $c_2$ only shifts the curve. Therefore, it is possible to use the shape of $Q_3^G$ in observations: when compared to the dark matter predictions  for $Q_3$ one can separate $b_1$ from $b_2$ in the above relation. This gives an estimate of the linear bias $b_1$ which is independent of the overall amplitude of clustering (eg. $\sigma_8$).  This approach has already 
been implemented for the skewness $S_3$ (Gazta\~naga 1994, Gazta\~naga \& Frieman 1994), the bispectrum (Frieman \& Gazta\~naga 1994, Fry 1994, Feldman et al. 2001, Verde et al. 2002) or the angular 3-point function (Frieman \& Gazta\~naga 1999).

We now test these ideas with mock galaxy catalogues.

\subsection{Mock galaxy samples}
\label{mocks}

We use three sets of mock catalogues, which we will called {\em 2dFGRS mocks}, {\em HV mocks} and {\em HOD mocks}. Each set has different charactertistics which can be used to test the different approximations used. 

\subsubsection{2dFGRS mocks}
 
The  2dFGRS mocks consist of 22 independent mock 2dFGRS surveys, extracted from the HV simulations (Evrad et al 2002) and  described in Norberg et al. (2002). They have the same radial and angular selection function as the final (public) version of the 2dFGRS and have been convolved with the completeness  mask of the survey (Cole etal 2005). We extracted 4 volume limited samples shown in Table~\ref{tab3}. These same samples have been used by Baugh et al.~(2004) and Croton et al.~(2004a,b) to study higher-order moments of counts-in-cells and void probabilities. Unless stated otherwise, we will use as a default the volume limited mock with 
$-21<M_b<-20$, which yields the best constraints (see Fig.\ref{b1c2gal}).

\begin{table*}
  \centering
  \footnotesize
  \caption[]{Properties of the combined 2dFGRS SGP and NGP volume limited
  catalogues (VLCs). Columns 1 and 2 give the faint and bright
  absolute magnitude limits that define the sample. Column 3 gives the
  median magnitude of the sample, computed using the Schechter
  function parameters quoted by Norberg et al. (2002).  Columns 4, 5
  and 6 give the number of galaxies, the mean number density and the
  mean inter-galaxy separation for each VLC, respectively.  Columns 7
  and 8 state the redshift boundaries of each sample for the nominal
  apparent magnitude limits of the survey; columns 9 and 10 give the
  corresponding comoving distances. Finally, column 11 gives the
  combined SGP and NGP volume.  All distances are comoving and are
  calculated assuming standard \LCDM.}
  \begin{tabular}{cccrccrrrrc} 
    \hline \hline
    \multicolumn{2}{c}{Mag. range} & Median mag. &  
           \multicolumn{1}{c}{N$_{\rm G}$} &
           {$n_{g}$} & {d$_{mean}$} & {z$_{min}$} & {z$_{max}$} &
           {D$_{min}$} & {D$_{max}$} & {Volume}\\
    \multicolumn{2}{c}{\tiny $M_{b_{\rm J}}-5\log_{10}h$} & 
           {\tiny $M_{b_{\rm J}}-5\log_{10}h$}& & 
           {\tiny $h^{3}/Mpc^3$} & {\tiny $h^{-1}$Mpc} & & &
           {\tiny $h^{-1}$Mpc} & {\tiny $h^{-1}$Mpc} & 
           {\tiny $10^6h^{-3}$Mpc$^3$}\\
           \hline
    $-$18.0 & $-$19.0 & $-$18.44 & 23290 & 9.26 & 0.0048 & 0.014 & 0.088 & 39.0 & 255.6 & 2.52\\
    $-$19.0 & $-$20.0 & $-$19.39 & 44931 & 5.64 & 0.0056 & 0.021 & 0.130 & 61.1 & 375.6 & 7.97\\
    $-$20.0 & $-$21.0 & $-$20.28 & 33997 & 1.46 & 0.0088 & 0.033 & 0.188 & 95.1 & 537.2 & 23.3\\
    $-$21.0 & $-$22.0 & $-$21.16 &  6895 & 0.110 & 0.0209 & 0.050  & 0.266 & 146.4 & 747.9 & 62.8\\
   \hline \hline
  \end{tabular}
  \label{tab3}
\end{table*}  

\subsubsection{HV mocks}

These consist of a set of 100 cubic subvolumes of side $L=350~$Mpc/$h$, extracted directly from the HV simulation (see Table~\ref{tab1}). There is no galaxy biasing, dilution or realistic mask involved in these mock catalogs. We will use them to estimate close to ideal conditions in a dark matter distribution. The HV mocks each have 2-3 times more effective volume than the 2dFGRS mocks.
 
\subsubsection{HOD mocks}
 
We also build galaxy mock catalogs following the results from the measurements of the two-point function in the SDSS (Zehavi et al. 2004).  For the thresholds $M_r<-20$ and $M_r<-21$, 
we use the VLS simulation, and a halo occupation distribution (HOD) prescription for the mean number of galaxies in a halo of mass $m>m_{\rm min}$

\beq
\langle N_{\rm gal}(m)\rangle = 1+\Big(\frac{m}{m_1}\Big)^\alpha \ \exp [-m_s/(m-m_{\rm min})],
\label{HOD}
\eeq
and zero otherwise, where the first contribution is that due to a central galaxy, the rest being satellite galaxies which are taken with a Poisson distributed scatter (Kravtsov et al. 2003). The power-law index  $\alpha$ is fixed to 1. For $M_r<-19$ and $M_r<-18$ mock galaxies were kindly provided by A.~Berlind, based  on the GIF simulation, with the same HOD as in Eq.~(\ref{HOD}) except for a sharp cutoff at $m_{\rm min}$ instead of an exponential cutoff, and the power-law index $\alpha$ is allowed to vary. 
The parameters of the HOD's are given in Table~\ref{tab2}. 

\begin{table*}
\begin{tabular}{|c|c|c|c|c|c|clcl} \hline
& $M_{\rm min}$ & $m_1$  & $m_s$ &  $\alpha$ &  $n_g$ & $b_1$ & $c_2$ \\ \hline
\hline
$M_r<-18$&$1.87\times 10^{11}$&$3.74\times 10^{12}$&$-$&0.92& 0.026&0.97&-0.21\\ 
$M_r<-19$&$3.87\times 10^{11}$&$8.8\times 10^{12}$&$-$&1.08& 0.013 &1.07&-0.08\\ 
$M_r<-20$&$9.9\times 10^{11}$&$1.71\times 10^{13}$&$6.86\times 10^{12}$&1& 0.005 &1.13&-0.11\\ 
$M_r<-21$&$5.09\times 10^{12}$&$5.77\times 10^{13}$&$5.77\times  10^{13}$&1&0.001&1.32&-0.03\\ \hline
\end{tabular}
\caption[junk]{Parameters of the mean halo occupation distribution used to populate dark matter halos in simulations with mock galaxies, see Eq.~(\protect\ref{HOD}). All masses are in units of $M_{\sun}/h$. The last three columns correspond to the number density $n_g$ (in galaxies per Mpc$^3h^{-3}$) and bias parameters calculated from Eq.~(\protect\ref{bhod}), with $c_2\equiv b_2/b_1$.}
\label{tab2}
\end{table*}

Given a HOD, the expected large-scale bias parameters (in the limit of large scales) are given by,

\beq
b_i = \frac{1}{n_g} \int dm\ n(m)\ \langle N_{\rm gal}(m) \rangle\ b_i(m),
\label{bhod}
\eeq

\noindent where $n(m)$ is the halo mass function (which we take to be that of Sheth and Tormen 2002), $b_i(m)$ are the corresponding halo bias parameters (Scoccimarro et al. 2001) and the galaxy number density is given by

\beq
n_g = \int dm\ n(m)\ \langle N_{\rm gal}(m) \rangle .
\label{ngal}
\eeq

Note how the different absolute magnitude limited samples have different
biasing parameters, contrary to what happens in the 2dFGRS mocks, which have
a fixed biasing prescription for all magnitude ranges. The main reason for using the HOD mocks is that we know the values of $b_1,b_2$ that we input in our catalogues, and we can therefore test how well $b_i$ can be recovered.

\begin{figure}
\centering{
{\epsfxsize=8cm \epsfbox{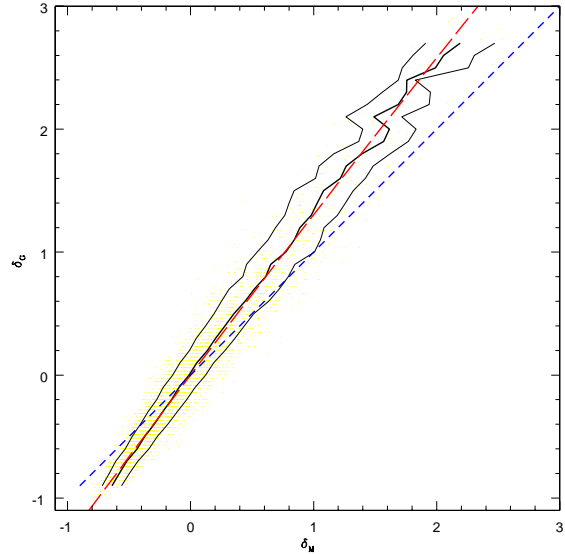}}}
\caption[junk]{Scatter plot showing galaxy density constrast $\delta_G$ versus dark matter $\delta_M$ fluctuations in the $M_r<-20$ HOD sample for cubic cells of side $l=30$ Mpc/$h$. The smooth long-dashed line shows the bias relation with parameters $b_1$ and $b_2$ given by the HOD from Eq.~(\protect\ref{bhod}) with parameters in Table~\ref{tab2}. The true mean bias relation is represented by the central thick line and 68\% scatter around it (parallel thin lines). The short-dashed line is the unbiased prediction, $b_1=1$, $b_2=0$.}
\label{deltas}
\end{figure}

Figure~\ref{deltas} shows how well these values of $b_1$ and $b_2$ fit the mean biasing scatter relation $\delta_G=F[\delta]$ in Eq.~(\ref{eq:bk}) for cubes of side $l=30$ Mpc/$h$ (see also Fig.~1 in Scoccimarro 2000 for more cases). As can be seen in the figure, the agreement is quite good. Similar results are found for other smoothing scales: $l>15$ Mpc/h. On smaller
scales, the scatter plot is dominated by shot-noise (the number of galaxies per pixel becomes too small).

\begin{figure*}
\centerline{\epsfysize=8truecm
\epsfbox{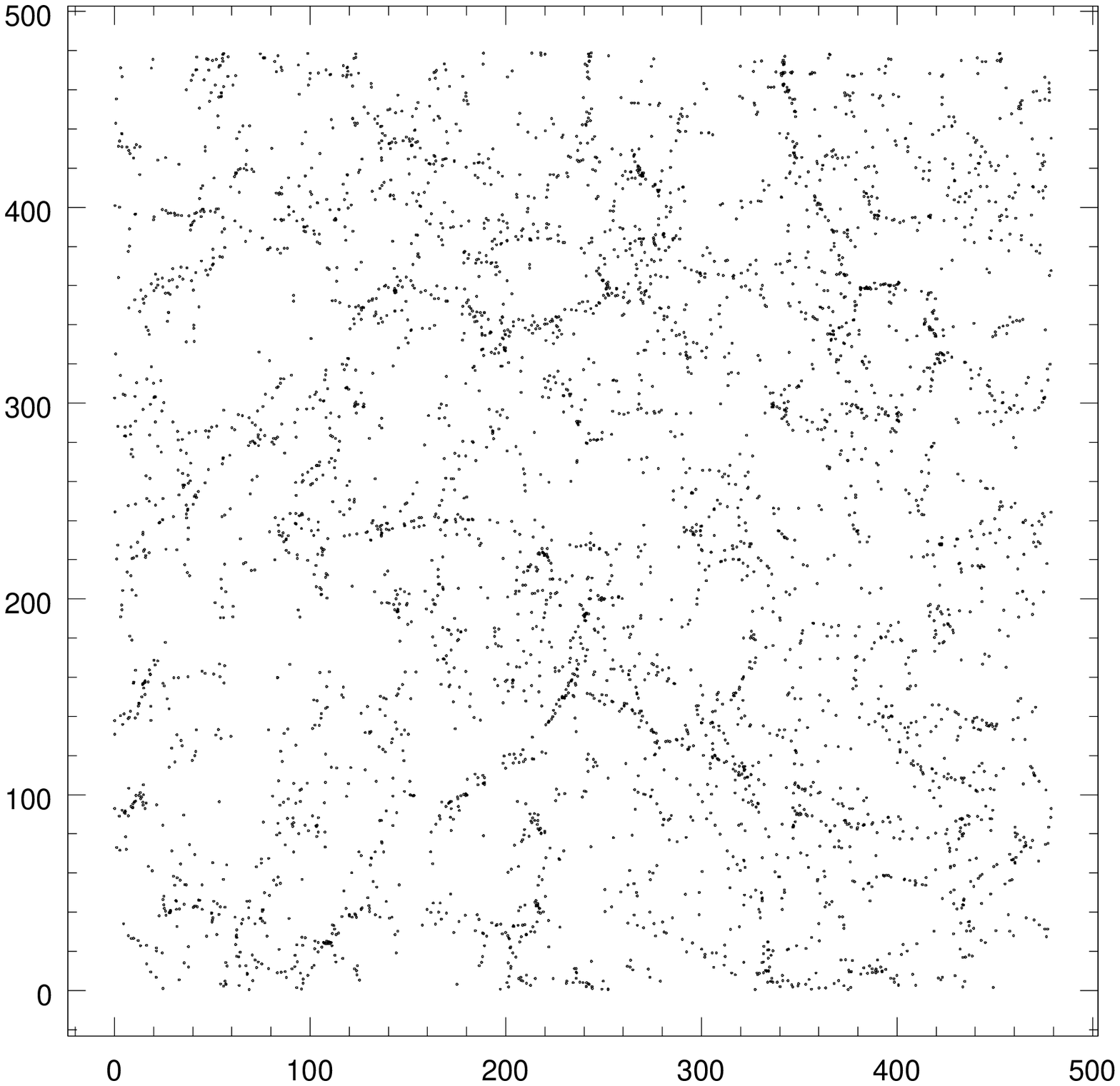}\epsfysize=8truecm
\epsfbox{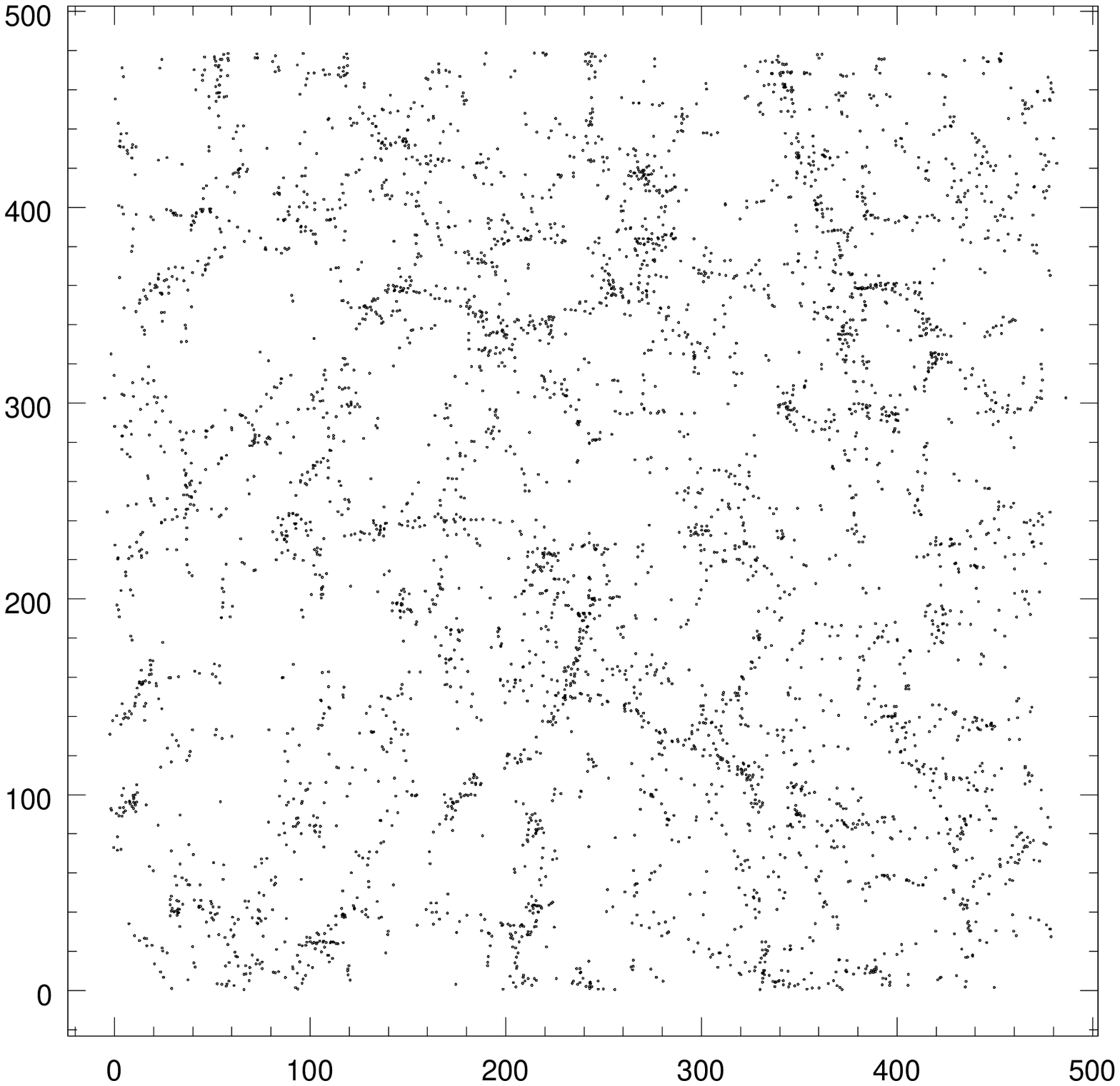}}
\caption[junk]{Galaxy positions in the real (left) and redshift space (right) in a 5 Mpc/$h$ slice corresponding to HOD  mock galaxies with $M_r<-20$. The observer is at $x \rightarrow -\infty$, just as in Fig.~\protect\ref{plotslice}.}
\label{galslice}
\end{figure*}

The HOD mocks do not have a realistic survey mask, as in the 2dFGRS mocks, but instead correspond to a simple square box geometry, as in the HV mocks. Given the large size of the VLS simulation 
(about 3 times in volume the HV mocks) and its simple  geometry, a single mock can be used as a prediction for the theoretically expected correlation functions for each set of bias parameters. The 22 2dFGRS and 100 HV mocks will be used to estimate errors and covariance matrices (see \S4). 

Figure~\ref{galslice}  shows a slice corresponding to the HOD mock galaxies with $M_r<-20$. Because of the low density we have used a thicker slice here than in the dark matter case of Fig.~\ref{plotslice}. Note how the fingers of god are not prominent in this case and is hardly possible to detect them visually (compare with  Fig.~\ref{plotslice}). We now study whether we can still detect in $Q_3$ the signature of fingers of god, the U-shape anisotropy. 
 
\subsection{Galaxy $Q_3$: Testing the validity of the bias expansion}

\begin{figure}
\centering{
{\epsfxsize=8cm \epsfbox{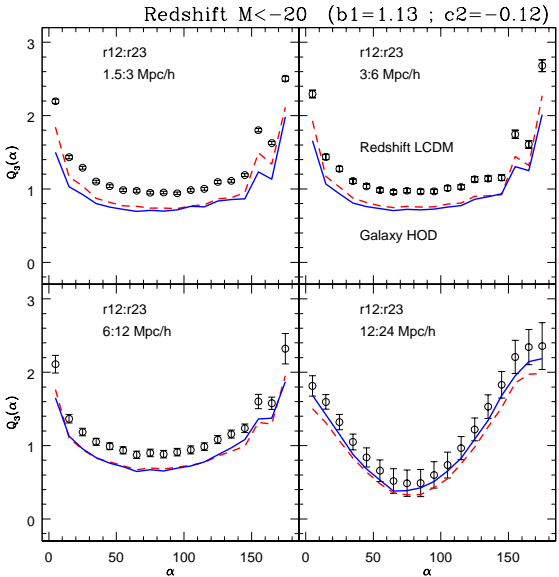}}}
\caption[junk]{
$Q_3$ in redshift space dark matter simulations (points with errorbars, from  100 HV mocks) and respective results for $M_r<-20$ HOD mocks (continuous line). Dashed lines shows the
HOD bias prediction using the values from Table~\ref{tab2} in Eq.~(\ref{eq:Q3G}).}
\label{q3gal}
\end{figure}

Figure~\ref{q3gal}  illustrates how  biasing changes the values of $Q_3$ from the dark matter simulations  (symbols with errors) to the HOD mock galaxies (continuous lines); all quantities here are in redshift space. The errors in the dark matter simulation are obtained from  the 100 HV mocks.  Similar results are found for other magnitude ranges. Note that the effects caused by biasing at the smallest scales are comparable to the variations among different models shown in Fig.~\ref{q3zr1}. The characteristic U-shape anisotropy due to fingers of god is also present in the galaxy samples on small scales $r_{12} \le 6 \mpc$, even though it is not so obvious to the eye in Fig.~\ref{galslice}. 

The HOD prediction based on Eqs.~(\ref{bhod}) and~(\ref{eq:Q3G}) (dashed lines) seems to work very well, even approaching the non-linear regime the deviations are not too large. Note however that the error bars become very small so these deviations are statistically significant at small scales, as we shall see in more detail below (see eg.  bottom panel in Fig.~\ref{b1c2}).  On the other hand, it is encouraging that the validity of the bias expansion seems to be much broader than perhaps naively expected, at least for the galaxy HOD's we have studied here. In this section and below we assume we can trust the local bias expansion, more precisely Eq.~(\ref{eq:Q3G}) for quantities in redshift space, where we always use simulations to make predictions for $Q_3$ for dark matter in redshift space. 

To characterize how well we can recover the actual bias parameters predicted by a given HOD it is necessary as we discuss now to study in detail the covariance between measurements of $Q_3$ for different triangle shapes and scales.
 
 \subsection{Q-eigenmodes}
 \label{Qeigen}
 
As we shall see in the next section, it is reasonable to expect the likelihood function to be Gaussian in terms of  $Q_3$, at least as a first approximation in the case of large surveys, which means that one can estimate parameters using a $\chi^2$ approach (likelihood ${\cal L} \propto \exp -\chi^2/2$), 
\beq
\chi^2 = \sum_{i=1}^{i=N_b} \sum_{j=1}^{j=N_b} \Delta_i C_{ij}^{-1} \Delta_j 
\label{eq:chi2}
\eeq
with $C_{ij}$ the normalized (to unit variance) covariance matrix and
$\Delta_i \equiv [Q_3^{obs}(i)-Q_3^{\rm mod}(i)]/\sigma_Q(i)$,
where $Q_3^{\rm mod}$ corresponds to the model (i.e. Eq.~(\ref{eq:Q3G}) with simulations used to predict the dark matter $Q_3$), $Q_3^{obs}$ is the measured reduced three-point function of the galaxy population whose bias parameters we want to determine, and $\sigma_Q$ are the uncertainties in $Q_3$ which are typically obtained from mock galaxy catalogs (as the diagonal part of the covariance matrix), or estimated directly from the observations eg. using independent sub-samples or
 jacknife resampling. The label  $i=1,\dots, N_b$ denotes different  triangle  shape configurations. In our case,  without any loss
of generality, we have fixed $r_{13}=2r_{12}$ and  bins refer only to changes in $\alpha$ with $N_b=18$ (i.e. $\Delta \alpha = 5$ deg.). In general one could mix many different shape configurations 
(i.e. different scales) and consider a larger covariance matrix (Scoccimarro 2000).

\begin{figure}
\centering{
{\epsfxsize=6.6cm \epsfbox{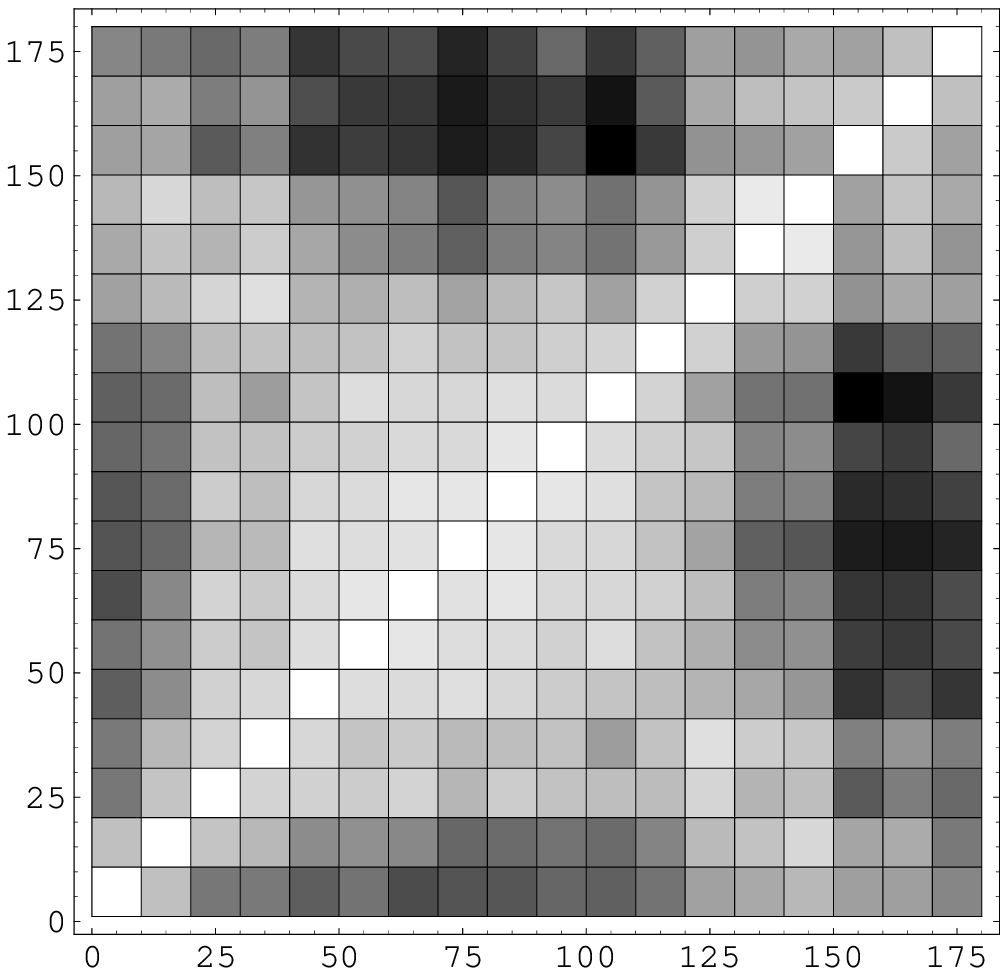}}
{\epsfxsize=6.6cm \epsfbox{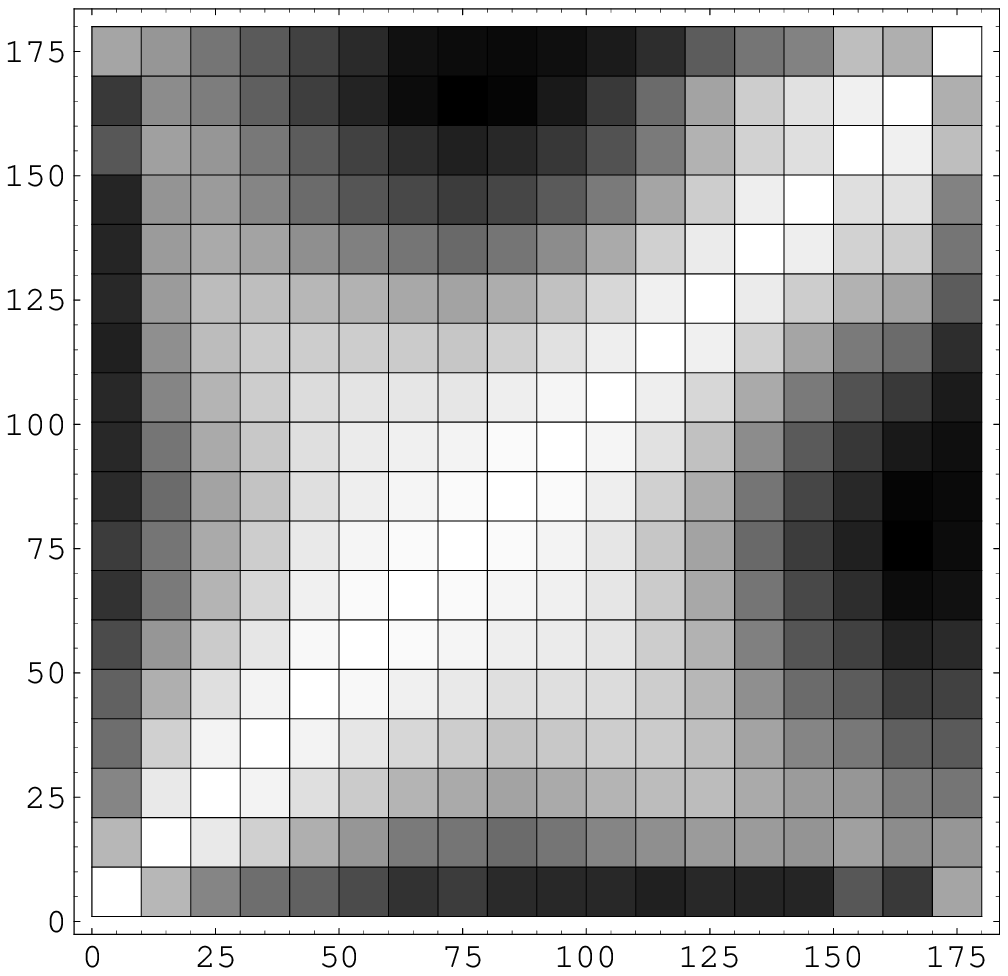}}}
\caption[junk]{Normalized covariance matrix $C_{ij}$ as a function of the configuration angle $\alpha$ 
 (in degrees) from $Q_3(\alpha)$ for fixed side triangles with $r_{13}=2r_{12}=24~$Mpc/$h$. White and black colours corresponds to $+1$ and $-1$ covariance, respectively. Middle grey corresponds to zero values. The top panel shows the covariance matrix from 22 2dFGRS mocks, while the bottom panel is obtained from the 100 HV mocks.}
\label{covQ3}
\end{figure}

 The normalized covariance matrix is estimated from the mock catalogs as, 
\beq
C_{ij} \equiv {1\over{N_m}} ~\sum_{k=1}^{k=N_m}~ \Delta_i^k \Delta_j^k
\eeq
with $\Delta_i^k \equiv [Q_3^k(i)-\bar{Q}_3(i)]/\sigma_Q(i)$, where $\bar{Q}_3(i)$ is the mean of the $N_m$ mock independent realizations denoted by the index $k$. Figure~\ref{covQ3} shows the estimated covariance matrices as a function of the configuration angle $\alpha$ in triangles of scale $r_{13}=2r_{12}=24~$Mpc/$h$.  As can be seen in this figure, the covariance is quite significant and cannot be neglected. The structure is very similar between the 2dFGRS (top) and HV (bottom) cases.

Before we invert $C_{ij}$ in Eq.~(\ref{eq:chi2}), notice that the values of $C_{ij}$  are estimated in practice to within a limited resolution, 
\beq
\Delta C_{ij} \simeq \sqrt{2\over{N_m}}, 
\eeq
therefore if $N_m$ is small or if there are degeneracies in $C_{ij}$ the inversion will be affected by numerical instabilities. In order to eliminate this problem, we perform a Singular Value Decomposition (SVD) of the matrix,
\beq
C_{ij}=(U_{ik})^\dagger D_{kl}V_{lj},
\label{eq:svd}
\eeq
where $D_{ij}= \lambda_i^2 \delta_{ij}$ is a diagonal matrix with the singular values on the diagonal, and $U$ and $V$ are orthogonal matrices that span the range and nullspace of $C_{ij}$ (see 
 Eisenstein \&  Zaldarriaga 2001, for an application to the power spectrum estimation). 
  By doing the SVD decomposition, we can choose the number of modes  we wish to include in our $\chi^2$ by effectively setting the corresponding inverses of the small singular values to zero. In practice, we work only with the subspace of ``dominant modes"  which satisfy:
\beq  
\lambda_i^2> \sqrt{2/N_m}
\label{eq:lambda}
\eeq
which is the resolution to which we can estimate the covariance matrix elements. In other words, when doing the inverse of $C$ in Eq.~(\ref{eq:chi2}), one inverts Eq.~(\ref{eq:svd}) with a restriction on $D_{kl}$ based on the criterion given by Eq.~(\ref{eq:lambda}).

\begin{figure}
\centering{
{\epsfxsize=6.6cm \epsfbox{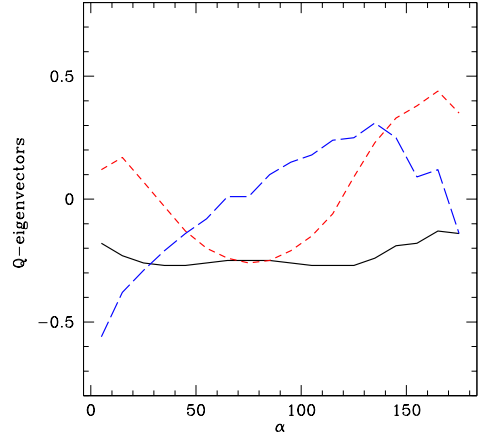}}}
\caption[junk]{The three dominant, in terms of signal to noise [see Eq.~(\ref{StoN}] Q-eigenvectors for
triangles with $r_{13}=2r_{12}=24$~Mpc/$h$ from 100 HV mock catalogs. The signal-to-noise values are $9.1$ (solid), $8.4$ (short-dashed), and $8.0$ (long-dashed).}
\label{eigen}
\end{figure}

Within the Gaussian assumption one needs to go no further than what we have discussed to obtain constraints on bias parameters. However, Gaussianity must be checked for each particular survey and scale under study.
 As shown for the bispectrum (Scoccimarro 2000), the non-Gaussian case can be simplified by rotating into the basis where the $Q_3$'s become diagonal, and their non-Gaussian distributions can be obtained from mock catalogs and  multiplied together effectively assuming that they are independent to construct the likelihood function for the bias parameters. 

In other words, one can represent each of the mock or data $Q_3$ vectors (the vector index $i=1,N_b=18$ runs over different values of $\alpha$ in this case) in a new basis of Q-eigenvectors where the covariance matrix becomes diagonal (since $U$ is the  rotation matrix
for nonzero modes ),
\beq
\widehat{Q}_3(i) = \sum_j U_{ji}\, \frac{Q_3(j)}{\sigma_Q(j)},
\label{eigenvec}
\eeq
The data expressed in the new basis, the {\em Q-eigenmodes} $\widehat{Q}_3$, is a vector which in general has a lower dimensionality than $N_b$ due to the criterion given by Eq.~(\ref{eq:lambda}).  Contrary to $Q_3$, the Q-eigenmodes are effectively statistically independent (this becomes rigorous in the Gaussian case), in the sense that their covariance matrix is diagonal. The eigenvalues $\lambda_i^2$ are the diagonal elements of their covariance matrix, i.e. the variance of a given Q-eigenmode. One can construct a ``signal to noise"  (S/N) for each eigenmode labeled by $i$, 

\beq
\Big(\frac{S}{N}\Big)_i =  \Big| \frac{\widehat{Q}_3(i)}{\lambda_i} \Big|= 
 \Big| \frac{1}{\lambda_i}\sum_{j=1}^{N_b} U_{ji} \, \frac{Q_3(j)}{\sigma_Q(j)} \Big|.
\label{StoN}
\eeq
The total  S/N can be obtained by adding the individual modes in quadrature. When we  use only the 
dominant   Q-eigenmodes,  the total S/N estimate corresponds to a lower bound, which is not
optimal, but avoids potential errors of a singular inversion.
Figure~\ref{eigen} shows the three most important (in terms of signal to noise) Q-eigenvectors for the case of the HV mocks at scales $r_{13}=2r_{12}=24~$Mpc/$h$. As can be seen in the figure the Q-eigenvectors correspond to interesting linear combination of different triangles: the first mode (solid) is basically the average $Q_3$ (slightly down-weighted for collapsed configurations, which have larger errors), the second eigenmode (short-dashed) measures the difference between collapsed and perpendicular configurations, this is sensitive to the dependence of $Q_3$ on triangle shape, the third eigenvector (long-dashed) gives the difference between the configuration dependence at small and large $\alpha$ (down-weighted on the largest scales
because of the large errors), this is sensitive to the scale dependence of $Q_3$. The behavior of the eigenvectors shown in Fig.~\ref{eigen} is completely analogous to what was found in the bispectrum case (see Fig.~16 in Scoccimarro 2000). The condition in Eq.~(\ref{eq:lambda}) is satisfied by 6 Q-eigenmodes 
(individual signal-to-noise values are $S/N=9.1,8.4,8.0,6.7,4.9$ and $1.1$)
in this particular case with a total signal-to-noise of $S/N=17.0$. 
 Later, in section~\ref{recovBias} and Figs.~\ref{b1c2}-\ref{b1c2gal}, we will illustrate the use of the Q-eigenmodes to recover the bias parameters. 

\begin{figure*}

\centerline{\epsfysize=5truecm
\epsfbox{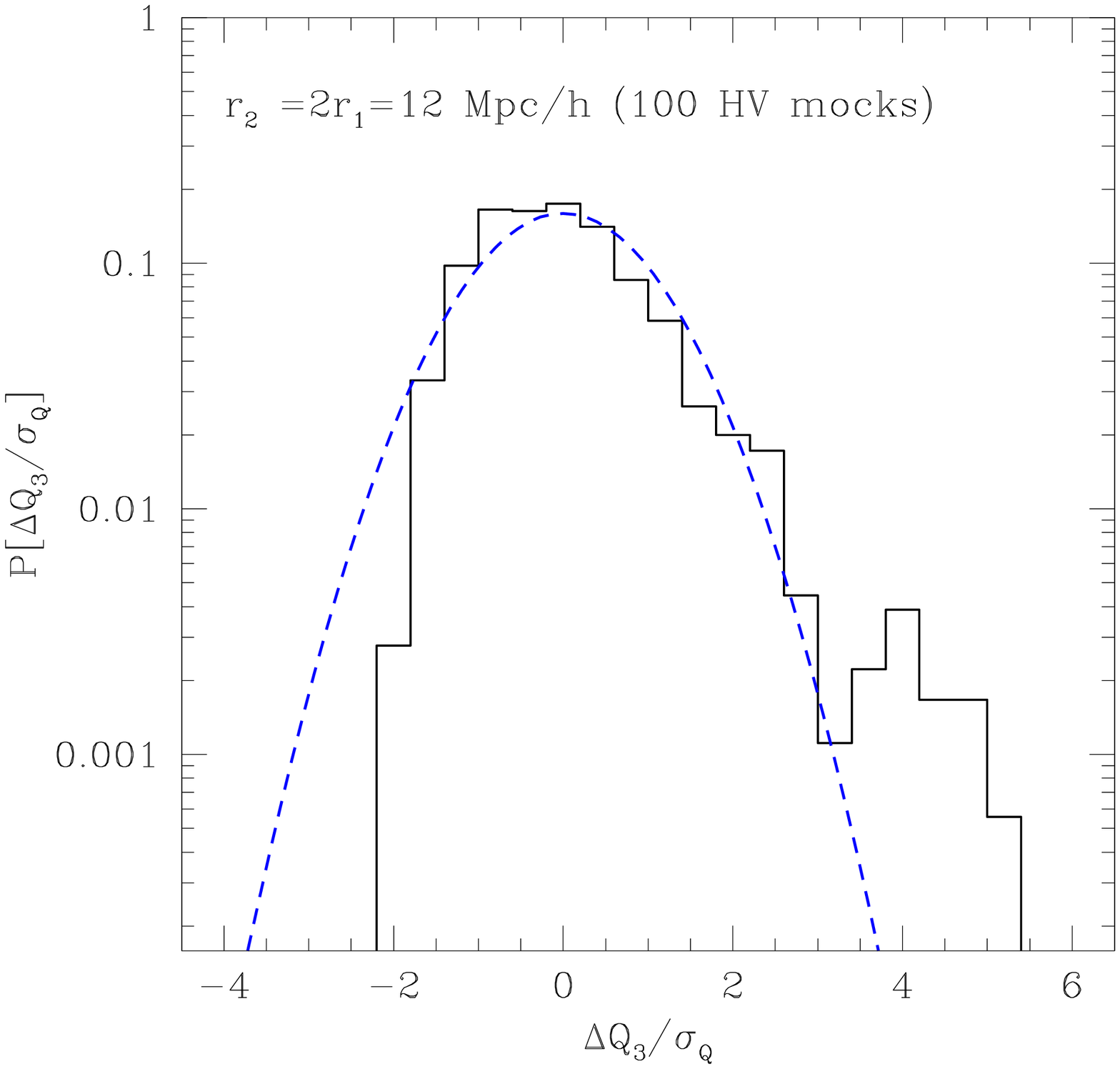}\epsfysize=5truecm
\epsfbox{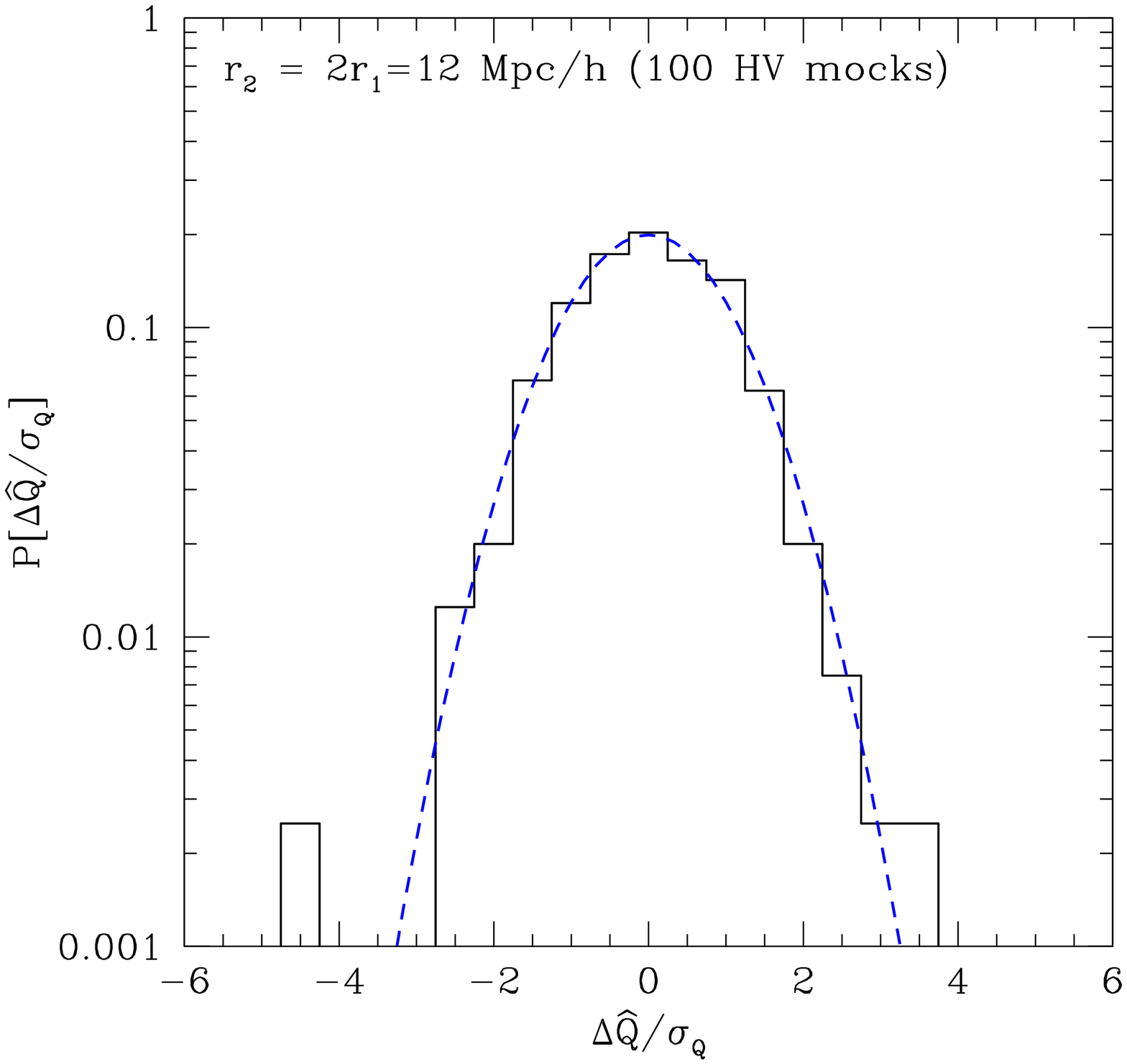}\epsfysize=5truecm
\epsfbox{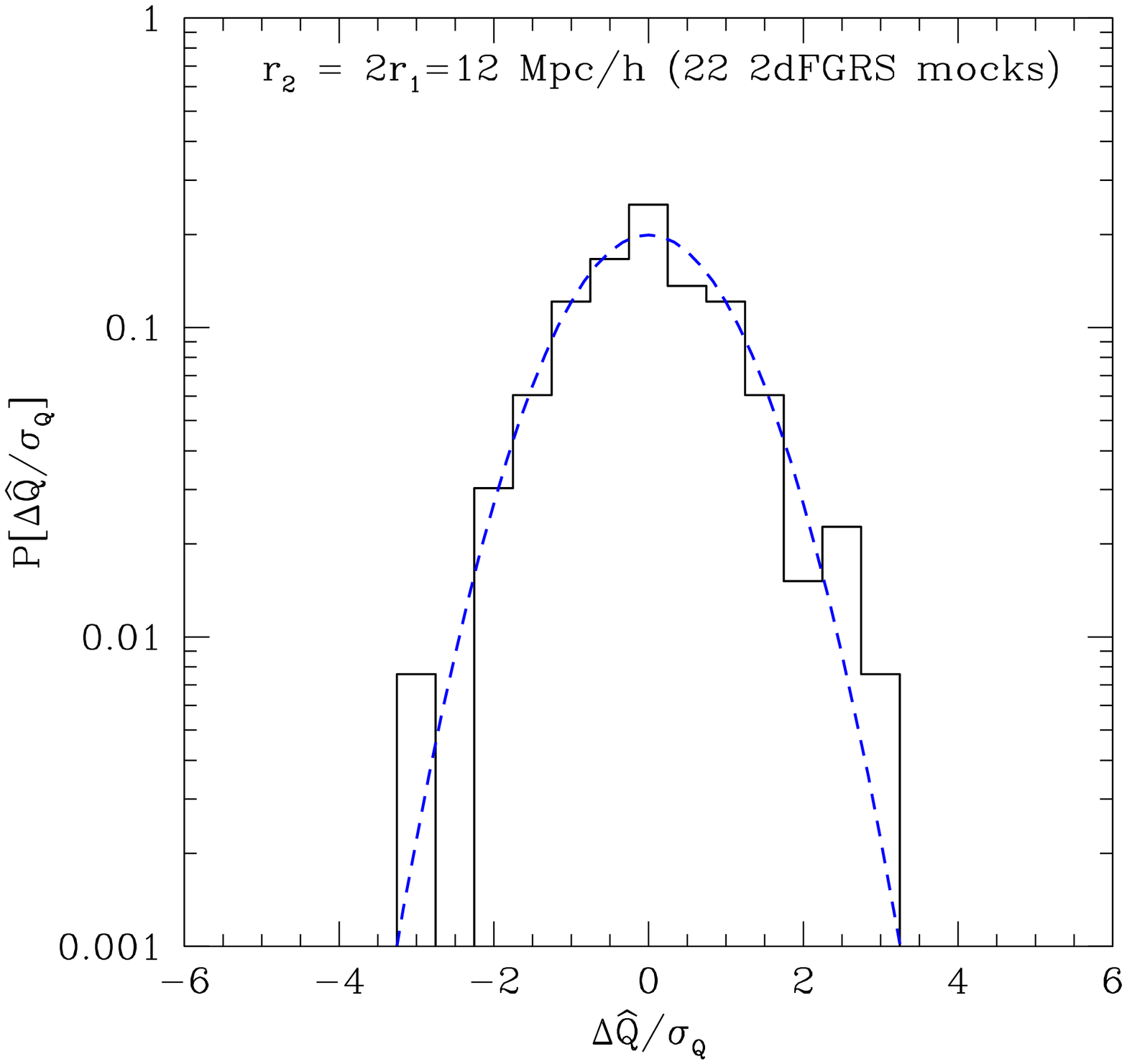}}
\caption[junk]{
Histograms comparing the measured distribution of 3-point amplitudes $Q_3$ (normalized to have unit variance) to the expected Gaussian distribution.  The left panel corresponds to the 100 HV mocks, 
which results in a slightly skewed non-Gaussian distribution. The middle panel shows the distribution for the 5 dominant Q-eigenmodes $\widehat{Q}$ using the same data, which is much more Gaussian.
The right panel shows the first 5 dominant Q-eigenmodes $\widehat{Q}$ in the 22 2dFGRS mocks.
All cases correspond to triangles with fixed sides $r_{12}=2r_{13}=12~$Mpc/$h$ and different values of $\alpha$.}
\label{pdq3}
\end{figure*}

\subsection{Gaussianity of the Likelihood Function}
\label{GLike}

A central assumption we made above is that the Q-eigenmodes are Gaussian distributed. This can {\em and should always} be checked against numerical realizations of the particular survey geometry and scales being studied, otherwise the outcome of the likelihood analysis could be significantly biased, even if  the covariance between different triangles is taken into account. Here we explore the validity of the Gaussian likelihood approximation for the three-point function, extending similar studies done for the bispectrum (Scoccimarro 2000) and higher-order moments (Szapudi et al. 2000).  
 
As the survey volume increases, for a triangle of given scale and shape, the Gaussian approximation should become better as more such triangles contribute to its $Q_3$ and although one is adding correlated quantities, one naively expects to reach the central limit theorem if the correlations are small. However, it is not obvious this is enough because at the same time the error bars for the bias parameters shrink, and thus a small systematic bias (if comparable to the small error bars) can still affect the determination of parameters. A simple rule of thumb is to compute the skewness of the distribution of $Q_3$ and check whether this is much smaller than the shift in $Q_3$ that results from displacing $b_1,b_2$ by the error bars, if so the systematic error that results from assuming Gaussianity can perhaps be safely ignored (see discussion in Hui \& Gazta\~naga 1999, Scoccimarro 2000
and Szapudi et al. 2000 for more details).
 
Here we compare the histograms of values of $Q_3$ in the galaxy mocks to a normalized Gaussian distribution: for each triangular shape, $\alpha$, $r_{12}$ and $r_{13}$, we estimate the mean
value of $Q_3$, $\bar{Q_3}$, and its variance $\sigma_Q^2$ in the mock catalogs. We then build a histogram of  $(Q_3 -\bar{Q_3})/\sigma_Q$ values for different realiazations and also different configuration angles $\alpha$, but  with fixed $r_{12}$ and $r_{13}$ sides. Although in general the distribution is not far from a Gaussian, it sometimes shows significant deviations.

The left panel in Fig.~\ref{pdq3} shows the worst situation we have encountered in our study. The distribution is somewhat skewed, although this is not a large effect (notice the logarthmic scale). This is much more reassuring than the analogous result for IRAS surveys (see e.g. Fig.~15 in Scoccimarro 2000) where the distribution  becomes strongly non-Gaussian for smaller galaxy samples. In our case, the 2dFRGS or HV mocks are large enough to avoid these severe biases and the Gaussian likelihood seems a good approximation, but the distribution still has some small degree of skewness. In  the left panel in Fig.~\ref{pdq3} 
 the dimensionless skewness of the distribution is $+1.1$, which translates in a bias of about
 $10\%$ in the estimated $Q_3$.

The relevant distribution in our method is not so much the distribution of $Q_3$, but the distribution of the Q-eigenmodes. These are shown in the middle (HV mocks) and right (2dFGRS mocks) panels of Fig.~\ref{pdq3} (notice the change in vertical scale from the left panel). The distribution seems better approximated by a Gaussian, which is expected since these are the highest signal to noise modes and should approach Gaussianity fastest. The results correspond to triangles with fixed sides $r_{12}=2r_{13}=12~$Mpc/$h$, but similar results are found for other scales.  The dimensionless skewness in both cases is of order
$0.1$, much smaller than in the direct distribution of $Q_3$. This translates into $1\%$ bias in
$Q_3$, which is smaller that the sampling variance from the single HOD mock.

We conclude that a Gaussian likelihood provides a good approximation for our particular situation, specially when we express our results in terms of Q-eigenmodes. We will therefore use the Gaussian 
$\chi^2$ test in Eq.~(\ref{eq:chi2}), restricted to the subspace of the dominant Q-eigenmodes from the SVD criterion,  Eq.~(\ref{eq:lambda})
 
\subsection{Recovering Bias Parameters}
 \label{recovBias}

\begin{figure*}
\centering{\vspace{-1.cm}
{\epsfxsize=12.5cm \epsfbox{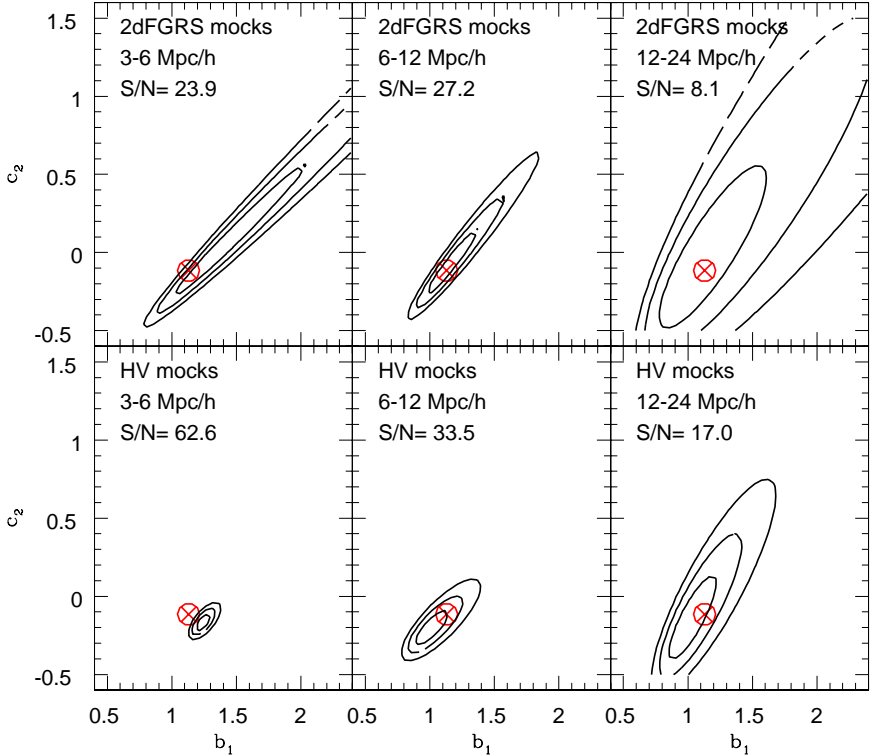}}
\vspace{-0.5cm}}
\caption[junk]{
Constraints on $b_1$ and $c_2$ for triangles of different scales, as labeled in each panel. In all cases we compare measurements of $Q_3$ in dark matter simulations to galaxy HOD mock catalogs. The top panels use errors (and Q-eigenmodes) from 2dFGRS mocks, while the bottom panels use HV mocks (smaller contours because HV have 2-3 times the effective volume of 2dFGRS mocks).}
\label{b1c2}
\end{figure*}

Figure~\ref{b1c2} shows how well we can recover the values of $b_1$ and $c_2$ using the measurements of $Q_3$ for  $r_{13}=2r_{12}=6, 12$ and $24 \mpc$ assuming that the  clustering 
pattern follows that of  the $M_r<-20$ HOD mock galaxies. We use two different sets of mock catalogs (HV and 2dFGRS) to estimate the error bars and Q-eigenmodes to assess the sensitivity of each survey geometry. The mean values for the dark matter and galaxy $Q_3$  are from the same simulation (VLS). From inside out, the values of the contours denote $\Delta \chi^2 \equiv \chi^2 - \chi^2_{min} = 2.4, 6.2, 11.8$ away from the minimum $\chi^2$ value in Eq.~(\ref{eq:chi2}), corresponding to $1$, $2$ and $3\sigma$ confidence intervals for a $\Delta \chi^2 $ distribution with two free parameters. Each panel also displays the  total S/N, according to Eq.[\ref{StoN}], added in quadrature for  the
dominant Q-eigenmodes used in each case.

As can be seen in Fig.~\ref{b1c2}, there is a main direction that is best determined corresponding to the first eigenmode, this is basically the direction perpendicular to the line $(Q_{\rm avg}+c_2)/b_1={\rm const.}$ where $Q_{\rm avg}$ is the average value of $Q_3$. As small scales are approached, $Q_3$ develops the characteristic U-shape anisotropy and if the errors do not decrease fast enough (top panel) a near degeneracy develops  in the $b_1-c_2$ plane. The reason is that  $b_1$ scales the shape dependence of $Q_3(\alpha)$, while $c_2$ only shifts the curve up or down. In the limit where the "U" is made of vertical and horizontal lines alone, scaling ($b_1$) and shifting ($c_2$) produce equivalent (degenerate) effects.
 For large scales, the contours become wide because of the lack of enough independent triangles. As a result of these two limits, the best configurations for constraining the $b_1-c_2$ parameter space in 2dFGRS are those at intermediate scales  ($r_{13}=2r_{12}=12 \mpc$). 

The crossed circle in  Fig.~\ref{b1c2} corresponds to the $b_1,c_2$ expected for the input galaxy mock catalogues using the HOD predictions in Eq.~(\ref{bhod}). These  predictions  are also shown as dashed lines in Fig.~\ref{q3gal} and provide a very good description of the mean bias relation, at least on large scales.  Note that there are no free parameters here, we are not fitting for $b_1$ and $c_2$, but rather checking we recover 
the values used for $b_1$ and $c_2$ to create the mock catalogs using parameters in Table~\ref{tab2}.
Also recall that we are using a single HOD mock (from VLS simulations) to test the predictions, which means that  there is some variance in the estimation and we should not expect the crossed circle on top of the best fit value.

On smaller scales, our treatment of biasing breaks down because one can no longer use the expansion in Eq.~(\ref{eq:bk}) nor the simple HOD predictions based on Eq.~(\ref{bhod}). This failure can be seen in the bottom left panel of Fig.~\ref{b1c2}, where one finds significant ($> 3\sigma$) deviations from the predicted values. 

\begin{figure}
\centering{
{\epsfxsize=7cm \epsfbox{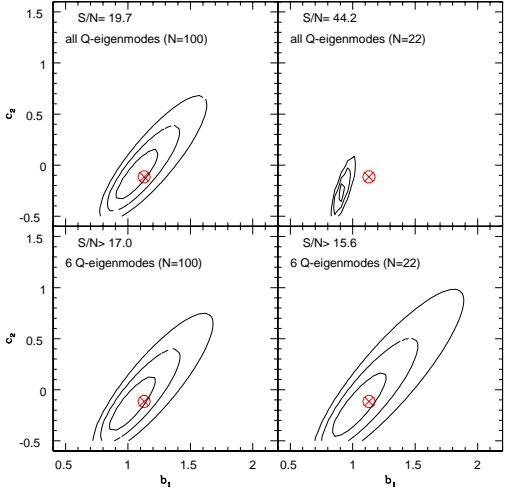}}}
\caption[junk]{
Constraints on $b_1$ and $c_2$ using the data in Fig.~\protect\ref{q3gal} ($M_r<-20$)
at $r_{13}=2r_{12}=24 \mpc$. The input values (in Table \ref{tab2}) are denoted with a crossed circle.  The top panels use all the Q-eigenmodes  (equivalent to a direct inversion of the covariance matrix), while the bottom panels use only the dominant Q-eigenmodes (Eq.[\ref{eq:lambda}]). Left panels use 100 HV mocks while right panels use only the first 22 HV mocks. Using the direct covariance inversion with too few mock catalogs (top right panel) typically gives systematically biased results and unrealistically small errorbars and large S/N.}
\label{b1c2r12hvM20all}
\end{figure}

Figure~\ref{b1c2r12hvM20all} illustrates the importance of using the dominant Q-eigenmodes, specially when we only have a small number of mock catalogs or samples to estimate the covariance matrix. As illustrated by the top right panel, if a noisy covariance matrix is inverted directly one recovers the incorrect values with unrealistically  small  error bars  due to numerical instabilities. When we restrict the number of Q-eigenmodes the errors  become larger (compare top and bottom left panels), but the
estimation is not biased by the number of mocks used (compare two bottom panels).  Note how the S/N is
wrongly estimated in the top right panel and corresponds to a lower estimate when we use
only the dominant   Q-eigenmodes .

\begin{figure}
\centering{
{\epsfxsize=7cm \epsfbox{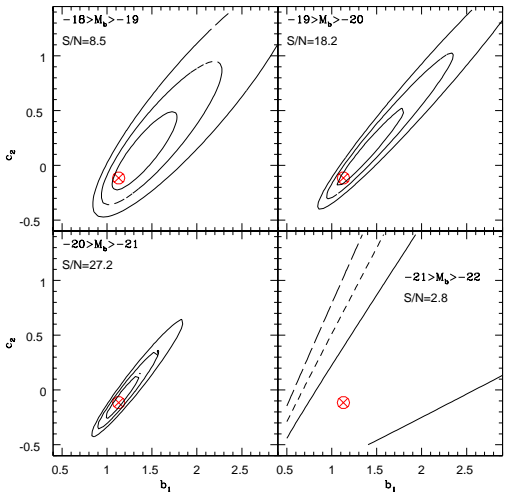}}}
\caption[junk]{
Constraints on $b_1$ and $c_2$ from realistic 2dFGRS mocks (using the dominant  Q-eigenmodes) of different absolute magnitudes (the volume-limited samples in Table~\ref{tab3}). Here we only use triangles with $r_{13}=2r_{12}=12 \mpc$.}
\label{b1c2gal}
\end{figure}

Finally, Fig.~\ref{b1c2gal} shows the constraints for different volume limited samples for the 2dFGRS mocks. The case  $-21<M_b<-20$  provides the best compromise between sample volume and number density. The brighter sample has more volume, but is dominated by shot-noise. The fainter samples
with higher density do not cover a large enough volume. These results are qualitatively similar to Fig.~9 in Sefusatti \& Scoccimarro (2004).

\section{Conclusions}

We have studied the three-point function amplitude $Q_3$ in dark matter and mock galaxies with particular emphasis on the effects of redshift-space distortions and galaxy biasing. 

Whereas the dependence of $Q_3$ on triangle shape is partialy erased in redshift space on large scales $r_{12} > 10 \mpc$, at smaller scales $Q_3$ develops a characteristic U-shape anisotropy due to the effects of velocity dispersion. This is  expected based on visual inspection of the distribution in redshift space (the finger of God  effects). But nevertheless, it has not been detected before in simulations nor observations. We studied why this is so and concluded that low resolution in the configuration angle $\alpha$ can suppress this sharp feature due to mixing of collapsed configurations (which are few) with the more numerous (with much less amplitude) configurations away from $\alpha=0,180$ deg. This problem can be exacerbated by using the $v$ parameter as a probe of angle (instead of $\alpha$), which has been a common choice in small-scale studies. Moreover,
errors are largest for the collapsed configurations, which both makes it more difficult to
detect and is subject to larger  estimation biases.

Our detection of the U-shape anisotropy is completely analogous to the results from the small-scale bispectrum, which shows a very similar effect (Scoccimarro et al. 1999), and we see it in every cosmological model and including galaxy bias (see Figs.~\ref{q3zr1},~\ref{q3gal}).  We therefore conclude that the redshift-space three-point function {\em does not} follow the hierarchical ansatz at small scales (less than 6 Mpc/h), much less so than in real-space, contrary to statements in the literature. On the other hand, the {\em scale} dependence of $Q_3$ is weakly  suppressed in redshift-space when compared to real-space, thus if the U-shape anisotropy of $Q_3$ is missed due to poor resolution, a hierarchical three-point function becomes a good model.

We presented a detailed method for obtaining constraints on galaxy bias parameters from measurements of $Q_3$ in current galaxy redshift surveys, based on the Q-eigenmode analysis developed for the bispectrum (Scoccimarro 2000). We showed that the covariance matrix of $Q_3$ has non negligible extra diagonal components and presented an SVD inversion method that can be stable even if too few mock catalogs are used for the computation of the covariance matrix, with the only drawback in this case being an overestimate of the error bars.  We studied the Gaussianity of the likelihood function and showed that for surveys such as 2dfGRS Gaussianity is a good approximation for the distribution of the Q-eigenmodes, although some small skewness is present. 

We discussed the sensitivity of $Q_3$ to bias parameters as a function of the scale included in the analysis, and presented results for how well we can expect to recover the linear and quadratic bias parameters in the case of 2dFGRS. We showed that our method recovers the bias parameters introduced into mock galaxies by an HOD prescription and is also able to handle potential systematics in case a smaller number than ideal of mock catalogs is used to estimate the covariance matrix.

\section*{Acknowledgments}

We thank Andreas Berlind for making available to us the SDSS mock catalogs corresponding to $M<-18$ and $M<-19$ and Naoki Yoshida for help with the VLS simulations. We also thank Carlton Baugh, 
 Darren Croton and Peder Norberg for making the 2dFGRS and HV mocks available and for comments
 to the manuscript.
 We acknowledge support from  the Spanish Ministerio de Ciencia i
Tecnologia, project AYA2002-00850 with EC-FEDER funding and grants NSF PHY-0101738 and NASA  NAG5-12100. R.~S. thanks the Kavli Institute for Cosmological Physics at the University of Chicago for hospitality during a sabbatical visit. The GIF, VIRGO, VLS and HV simulations in this paper were carried out by the Virgo Supercomputing Consortium using computer s based at the Computing Centre of the Max-Planck Society in Garching and at the Edinburgh parallel Computing Centre. The data are publicly available at http://www.mpa-garching.mpg.de/NumCos


\end{document}